\documentclass[a4paper,11pt]{article}
\usepackage{pos}
\newcommand{\figw}[2]{\includegraphics[width=#1\textwidth]{#2}}

\title{Relativistic dust grains: a new subject of research with orbital
fluorescence detectors}

\ShortTitle{Relativistic dust grains with orbital fluorescence detectors}

\author{B.A. Khrenov}
\author{N.N. Kalmykov}
\author*{P.A. Klimov}
\author{S.A. Sharakin}
\author{M.Yu. Zotov}

\affiliation{Skobeltsyn Institute of Nuclear Physics, Lomonosov
Moscow State University, Moscow 119991, Russia}

\emailAdd{bkhrenov@yandex.ru}
\emailAdd{pavel.klimov@gmail.com}

\abstract{TUS (Tracking Ultraviolet Set-up) was the world's first
orbital detector aimed at testing the principle of observing ultra-high
energy cosmic rays (UHECRs) with a space-based fluorescence telescope.
TUS was launched into orbit on 28th April 2016 as a part of the
scientific payload of the Lomonosov satellite, and its mission continued
for 1.5 years.  During this time, its exposure  reached
$\sim1550$~km$^2$~sr~yr  for primary energy $\gtrsim400$~EeV, and a
number of extensive air showers-like events were registered.  The shape
and kinematics of the signal in these events closely resembled those
expected from UHECRs but amplitudes of the signal and some other
features were in contradiction with this assumption.  A detailed
analysis of one of EAS-like events (TUS161003) revealed that a primary
cosmic ray would need to have an energy $\gtrsim1$~ZeV in order to
produce a light curve of the observed amplitude, which is incompatible
with the cosmic ray spectrum obtained with ground-based experiments.
More than this, the slant depth of the shower maximum be the signal
produced by a cosmic particle, was estimated as $\lesssim500$~g/cm$^2$,
which corresponds to cosmic rays around 1~PeV.  We present a preliminary
discussion of a hypothesis that the TUS161003 event and perhaps some
other bright EAS-like events could be produced by relativistic dust
grains, which were considered a possible component of the cosmic ray
flux beyond the GZK cut-off some time ago.}

\FullConference{37$^{\rm{th}}$ International Cosmic Ray Conference (ICRC 2021)\\
		July 12th -- 23rd, 2021\\
		Online -- Berlin, Germany}

\begin{document}
\maketitle

%_______________________________________________________________________
\section{The TUS detector}

A comprehensive description of the TUS telescope can be found
in~\cite{SSR2017,JCAP2017}.
Here we will briefly outline its main features.

From the very beginning TUS, was designed as a pathfinder for a much
more sophisticated KLYPVE experiment onboard the International Space
Station~\cite{2001AIPC..566...57K, 2001ICRC....2..831A}.  The main
components of TUS were a Fresnel mirror and a square-shaped $16\times16$
photodetector aligned to the focal surface of the mirror.  The mirror
had an area of nearly 2~m$^2$ and a 1.5~m focal distance.  The field of
view (FOV) of the telescope was $9^\circ\times9^\circ$, which covered an
area of approximately $80~\text{km}\times80~\text{km}$ at sea level.
Totally 256 Hamamatsu R1463 photomultiplier tubes (PMTs) were grouped in
a $16\times16$ square and formed the focal surface of the photodetector.
The FOV of each channel was around $5~\text{km}\times5~\text{km}$ on
ground. A glass UV filter was placed in front of every PMT to limit the
measured wavelength to the 300--400~nm range.  Light guides with square
entrance apertures and circular outputs were used to uniformly fill the
FOV.  The high-voltage system was aimed to adjust the sensitivity of
PMTs to the intensity of the incoming light and switch them off
completely on day sides of the orbit.

The TUS electronics could operate in four modes with different time
sampling windows. The main mode was intended for registering the fastest
processes in the atmosphere and had the time step of~0.8~$\mu$s. Every
record consisted of ADC codes written for all photodetector channels in
256 time steps with a total duration of 204.8~$\mu$s. We only discuss
events registered in this mode of operation.

TUS was launched into orbit on 28 April 2016 as a part of the scientific
payload of the Lomonosov satellite. The satellite had a sun-synchronous
orbit with an inclination of $97.3^\circ$, a height about 470--515~km
and a period of $\approx94$ minutes.  The TUS mission continued till 30
November 2017, and the total exposure was estimated to be
$\sim1550$~km$^2$~sr~yr at energies
$\gtrsim400$~EeV~\cite{Fenu2021}.\footnote{A high energy threshold in
the estimation is partially due to a shortcut of the high-voltage system
that took place on the first day of TUS operation. The accident resulted
in 51 PMTs burnt out and a decrease of sensitivities of other channels.}

%_______________________________________________________________________

\section{EAS-like event TUS161003}

The TUS161003 event was registered on 3 October 2016 in approximately
100~km south-east from Minneapolis, MN, USA~\cite{JCAP2020}. The Richard
J.~Dorer Memorial Hardwood State Forest was in the field of view of hit
pixels of TUS, and no potential sources of artificial UV light were
identified on ground.
The signal was registered in perfect observational conditions without
any noticeable clouds except some small low-altitude ones, and low
background illumination.

\begin{figure}[!ht]
	\centering
	\figw{.7}{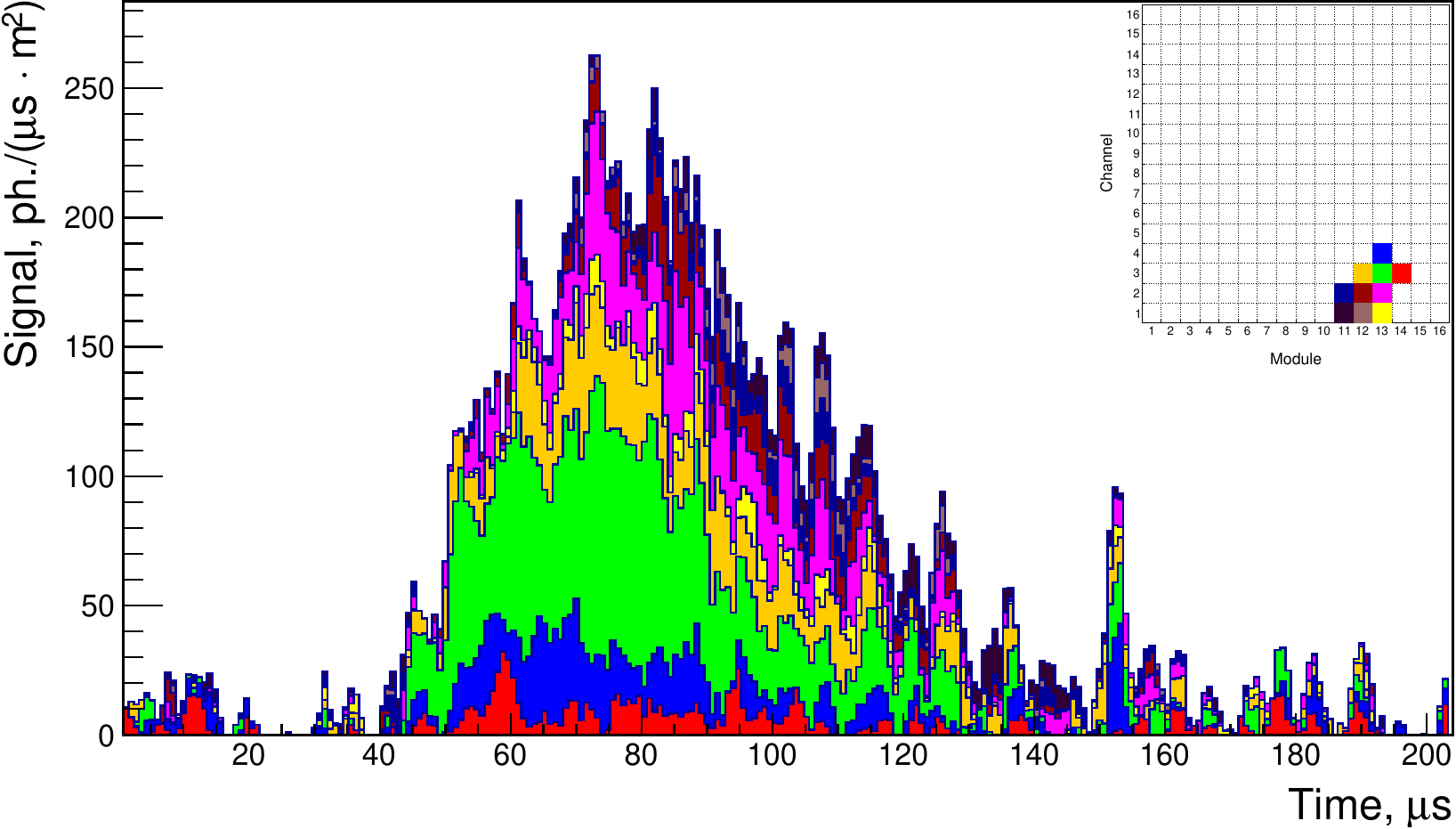}

	\caption{The light curve of the TUS161003 event and location of the
	hit channels in the focal surface of TUS~\cite{JCAP2020}.}

	\label{TUS161003}
\end{figure}

The light curve of the TUS161003 event and location of the hit channels
in the focal surface are shown in Figure~\ref{TUS161003}.  The
spatio-temporal dynamics of the signal found in 10 adjacent pixels was
similar to what was expected from an extensive air shower (EAS) basing
on detailed simulations performed with ESAF~\cite{ESAF}.  The arrival
direction of the signal source was found to be
$\theta=44^\circ\pm4^\circ$, $\phi=50^\circ\pm10^\circ$.  However, a
primary cosmic ray, be it a proton or an iron nucleus, would need an
energy $\gtrsim1~\text{ZeV}=1000~\text{EeV}$, which is at least three
times higher than the most powerful cosmic ray event ever registered
during more than 60 years of observations~\cite{1995ApJ...441..144B}.
The limited exposure of the TUS mission and results on the energy
spectra of the Pierre Auger Observatory and the Telescope
Array~\cite{2019ICRC...36..234D} make registration of such an extreme
energy event highly unlikely.

\begin{figure}[!ht]
	\centering
	\figw{.7}{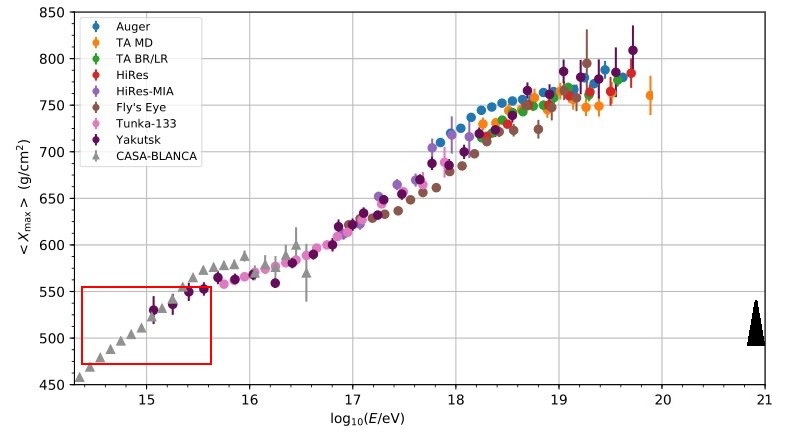}

	\caption{Mean depth of maximum of EASs vs.\ energy of primary cosmic
	rays according to data of different
	experiments. The figure is adapted from~\cite{TelescopeArray:2018xyi}.
	The red box shows the
	range of slant depths and respective energies estimated for the
	TUS161003 event. The black triangle shows estimations for the TUS161003
	event.}

	\label{spectrum}
\end{figure}

Another argument against interpreting the TUS161003 event as an UHECR
was the slant depth of the shower maximum. It was estimated from the
light curve as $\lesssim480-550$~g/cm$^2$, which geometrically
corresponds to altitudes $\sim7.5\text{--}8.5$~km above the ground.
This rules out a proton or an iron origin of the primary source of light
since a particle with an energy $\gtrsim1$~ZeV arriving at the zenith
angle $\sim44^\circ$ should produce and EAS that hits the ground before
reaching its maximum.  Conversely, the observed slant depth corresponds
to an EAS generated by a cosmic ray with an energy around PeV, see
Figure~\ref{spectrum}.

\section{Interpretation of TUS161003 as generated by a relativistic dust grain}

A number of artificial sources were considered in~\cite{JCAP2020}
in order to explain the TUS161003 event but only a peculiar
configuration of ground flashers was found to be able to reproduce
the kinematics and the light curve of the signal.
As was pointed out in~\cite{JCAP2020}, a possible way to explain the
TUS161003 event with an astrophysical phenomenon is provided by
relativistic dust grains (RDGs). They were first considered by
Spitzer~\cite{Spitzer1949} and later by Hayakawa~\cite{Hayakawa1972}
as possible sources of UHECRs of the highest energies.
The idea was revisited in~\cite{PhysRevD.61.087302, 2001NuPhS..97..203A}
and~\cite{2001AIPC..566...57K} and a number of other studies.
In particular, L.~Anchordoqui performed detailed simulations of EASs
produced by RDGs~\cite{PhysRevD.61.087302, 2001NuPhS..97..203A}
by means of the AIRES code\footnote{\url{http://aires.fisica.unlp.edu.ar/}}.
The simulations were based on a number of assumptions:

\begin{itemize}

	\item Relativistic dust grains encountering the atmosphere will
		produce a composite nuclear cascade.

	\item Each grain evaporates at an altitude of about 100~km and
		forms a shower of nuclei which in turn produces many small
		showers spreading over a radius of several tens of meters,
		whose superposition is observed as an extensive air
		shower.

	\item For the Lorentz factor $\Gamma \gg 1$, the internal forces
		between the atoms will be negligible.

	\item The nucleons in each incident nucleus will interact almost
		independently. Consequently, a shower produced by a dust grain
		containing~$N$ nucleons may be simulated by the collection
		of~$N$ nucleon showers, each with $1/N^{\rm th}$ of the grain
		energy.

\end{itemize}

Several sets of showers were generated in~\cite{PhysRevD.61.087302},
each one for with different Lorentz factors. The samples were distributed
in the energy range of $10^{18}$~eV up to $10^{20}$~eV and were equally
spread in the interval of 0$^{\circ}$ to 60$^{\circ}$ zenith angle at
the top of the atmosphere.  All shower particles with energies above
the following thresholds were tracked: 750~keV for gammas, 900~keV for
electrons and positrons, 10~MeV for muons, 60~MeV for mesons and 120~MeV
for nucleons and nuclei.  The particles were injected at the top of the
atmosphere (100~km a.s.l), and the surface detector array was put
beneath different atmospheric densities selected from the altitude of
cosmic ray observatories (Fly's Eye, Yakutsk, AGASA, Auger).
SIBYLL routines were used to generate hadronic interactions above
200~GeV.  Results of these simulations were compared with experimental
data, especially with an event recorded at the Yakutsk array on 7 May
1989~\cite{1991aame.conf..434E}.  One of the conclusions of the work was
that the dependence of the longitudinal profile of RDGs on the Lorentz
factor is rather weak, and while RDG air showers must be regarded as
highly speculative, they cannot be completely ruled out.

\begin{figure}[!ht]
	\centering
	\figw{.6}{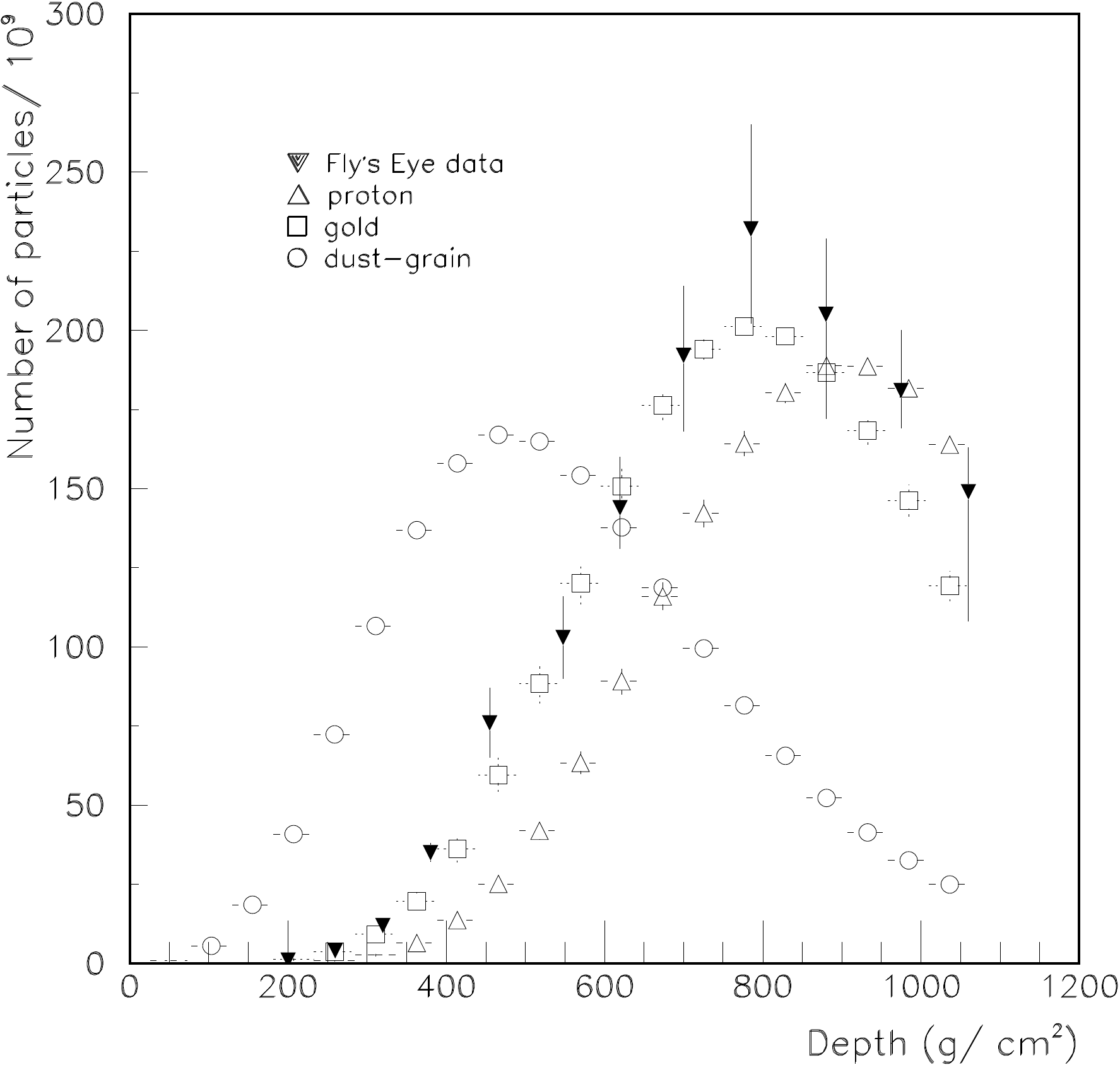}

	\caption{Longitudinal development of 300~EeV showers generated by a
	proton, a gold nucleus and a relativistic dust grain with
	$\lg\Gamma=4.5$ together with the data of the highest energy CR,
	registered by Fly's Eys. Figure from~\cite{2001NuPhS..97..203A}.}

	\label{anchor}
\end{figure}

The study was continued in~\cite{2001NuPhS..97..203A}. In this work,
Anchordoqui et~al.\ performed similar simulations but with a focus on
heavy and ``superheavy'' nuclei. In particular, they studied
longitudinal profiles of EAS generated by nuclei of different masses and
RDGs. The particles were injected vertically at the top of the
atmosphere, and the surface detector was put at sea level.
The energy of the highest energy event, registered by Fly's
Eye~\cite{1995ApJ...441..144B} was taken as a talon one.
One of their results is shown in
Figure~\ref{anchor}. Notice that the shape of the profile of an EAS
induced by a dust grain with the Lorentz factor $\lg\Gamma=4.5$
is similar to the light curve of the TUS161003 shown in
Figure~\ref{TUS161003}, and its maximum is located at the slant
depth $\approx500$~g/cm$^2$.

One of the key ingredients of the studies by Anchordoqui et~al.\ was the
superposition model, which assumes that an average shower produced by a
nucleus with energy~$E$ and mass number~$A$ is almost indistinguishable
from a superposition of~$A$ proton showers, each with energy~$E/A$. The
model is an approximate one but some experiments demonstrate that it
might be sufficiently solid.  In particular, a systematic study of meson
multiplicity as a function of energy in nucleus-nucleus collisions was
performed in~\cite{PhysRevLett.56.1350}, using cosmic-ray data in
nuclear emulsion.  It was shown that the data were consistent with the
superposition model.  The experiments gave identical results for final
nucleons energy distribution in two cases: when the full history of
primary nucleus fragmentation was taken into account, and when only
division of a primary nucleus with an energy~$E_0$ to~$A$ nucleons was
accounted.  The superposition model had worked if the absorption mean
free path in air of cluster fragments was much less than the absorption
mean free path of cluster nucleons.

Suppose we consider the atmosphere to be a target for a cluster of
nucleons (a dust grain) containing $N_\mathrm{n} = 10^6$ nucleons in
atomic and molecular states, with the primary energy of $E_0\approx
10^{21}$~eV and energy per nucleon $E_\mathrm{n} = E_0/N_\mathrm{n} =
10^{15}$~eV.  Applying the superposition model to this impact process,
one should expect as a final observable picture the sum of
$N_\mathrm{n}$ EASs with the total energy approximately equal to~$E_0$,
excluding the bond energy summed over all nucleons of the cluster.  For
a silicon nuclei with the number of nucleons in one nucleus $A=28$, the
bond energy per nucleon of about $\approx 8$~MeV and the number of
nucleons-grain constituents $N_\mathrm{n} =10^6$, the energy spent on
disintegration of all cluster constituents will be of the order of
$10^{14}$~eV, which is negligible in comparison with an RDG projectile
energy.

In case of applying the superposition model to RDG primaries, it is
important to keep in mind that molecules and atoms of RDG have very
small absorption mean free path in air
($10^{-5}\text{--}10^{-4}$~g/cm$^2$) in comparison with nuclear
absorption mean free path, which equals to 14~g/cm$^2$ for iron and
25~g/cm$^2$ for silicon, known as the  most abundant dust grains in
non-relativistic range. This implies that the atomic type of a cluster
impacting the atmosphere very soon (at altitudes of the order of a
hundred kilometers) changes their original form to the sum of
$N_\mathrm{n}$ independent nucleons.

\section{Discussion and Conclusions}

Orbital fluorescence telescopes aimed at observing UHECRs will provide
an interesting opportunity for studying relativistic dust grains.  While
the remarkable coincidence of the slant depth of a shower maximum
generated by an RDG shown in Figure~\ref{anchor} and that of the
TUS161003 event does not necessarily mean that the signal registered by
TUS could be produced by a dust grain, it suggests that this hypothesis
is worth studying in detail. Following the superposition principle
discussed above, one can check if an air shower generated by a
relativistic dust grain that consists of $\sim10^6$ protons with an
energy around 1~PeV is able to produce a light curve similar to that of
the TUS161003 event or some other EAS-like events registered by TUS.
Even in case it can, one should keep in mind that the superposition
model is only an approximation, and more advanced models might be
needed.  Besides this, other studies argue that photoelectronic emission
by optical and UV background radiation as well as Coulomb explosions due
to collisional charging in interstellar and intergalactic medium play an
important role in the destruction of relativistic dust grains in their
way in space, which considerably reduces chances that they account for
UHECRs~\cite{2015ApJ...806..255H}. This aspect should also be studied in
details.

Nevertheless, taking into account that statistics of events beyond the GZK
cut-off is very limited, and only a handful of UHECRs with energies
above 100~EeV have been registered~\cite{Auger}, we believe one cannot
completely rule out the possibility that a small fraction of cosmic rays
of the highest energies is produced by relativistic dust grains.  The
Mini-EUSO telescope~\cite{mini-EUSO} that currently operates at the
International Space Station and the future EUSO-SPB2~\cite{EUSO-SPB2},
K-EUSO~\cite{keuso} and the POEMMA~\cite{POEMMA:2020ykm} missions can
extend the capabilities of TUS and the ground-based detectors and shed
new light on this hypothesis.  Meanwhile, we are going to study this
possibility following the ideas outlined above and will report results
elsewhere.

\acknowledgments

The work was supported by Space State Corporation ROSCOSMOS and by
Lomonosov Moscow State University in frame of Interdisciplinary
Scientific and Educational School of Moscow University ``Fundamental and
Applied Space Research.''
%MZ thanks Luis Anchordoqui for an interesting discussion.

\newpage
\providecommand{\href}[2]{#2}\begingroup\raggedright\endgroup


\begin{thebibliography}{20}

\bibitem{SSR2017}
P.A.~{Klimov}, M.I.~{Panasyuk}, B.A.~{Khrenov}, G.K.~{Garipov},
  N.N.~{Kalmykov}, V.L.~{Petrov} et~al., \emph{The {TUS} detector of extreme
  energy cosmic rays on board the {Lomonosov} satellite},
  \href{https://doi.org/10.1007/s11214-017-0403-3}{\emph{Space Science Reviews}
  {\bfseries 212} (2017) 1687}
  [\href{https://arxiv.org/abs/1706.04976}{{\ttfamily 1706.04976}}].

\bibitem{JCAP2017}
B.A.~{Khrenov}, P.A.~{Klimov}, M.I.~{Panasyuk}, S.A.~{Sharakin},
  L.G.~{Tkachev}, M.Y.~{Zotov} et~al., \emph{First results from the {TUS}
  orbital detector in the extensive air shower mode},
  \href{https://doi.org/10.1088/1475-7516/2017/09/006}{\emph{Journal of
  Cosmology and Astroparticle Physics} {\bfseries 9} (2017) 006}
  [\href{https://arxiv.org/abs/1704.07704}{{\ttfamily 1704.07704}}].

\bibitem{2001AIPC..566...57K}
B.A.~{Khrenov}, M.I.~{Panasyuk}, V.V.~{Alexandrov}, D.I.~{Bugrov},
  A.~{Cordero}, G.K.~{Garipov} et~al., \emph{Space program {KOSMOTEPETL}
  (project {KLYPVE} and {TUS}) for the study of extremely high energy cosmic
  rays},  in \emph{Observing Ultrahigh Energy Cosmic Rays from Space and
  Earth}, H.~{Salazar}, L.~{Villasenor} and A.~{Zepeda}, eds., vol.~566 of
  \emph{American Institute of Physics Conference Series}, pp.~57--75, May,
  2001, \href{https://doi.org/10.1063/1.1378622}{DOI}.

\bibitem{2001ICRC....2..831A}
V.V.~{Alexandrov}, D.I.~{Bugrov}, G.K.~{Garipov}, V.M.~{Grebenyuk},
  M.~{Finger}, B.A.~{Khrenov} et~al., \emph{Space experiment {TUS} for study of
  ultra high energy cosmic rays}, {\emph{International Cosmic Ray Conference}
  {\bfseries 2} (2001) 831}.

\bibitem{Fenu2021}
F.~{Fenu}, K.~{Shinozaki}, M.~{Zotov}, M.~{Bertaina} and P.~{Klimov},
  \emph{Estimation of the exposure of the {TUS} space-based cosmic ray
  observatory}, these proceedings, 2021.

\bibitem{JCAP2020}
B.A.~{Khrenov}, G.K.~{Garipov}, M.A.~{Kaznacheeva}, P.A.~{Klimov},
  M.I.~{Panasyuk}, V.L.~{Petrov} et~al., \emph{An extensive-air-shower-like
  event registered with the {TUS} orbital detector},
  \href{https://doi.org/10.1088/1475-7516/2020/03/033}{\emph{Journal of
  Cosmology and Astroparticle Physics} {\bfseries 2020} (2020) 033}
  [\href{https://arxiv.org/abs/1907.06028}{{\ttfamily 1907.06028}}].

\bibitem{ESAF}
C.~{Berat}, S.~{Bottai}, D.~{De Marco}, S.~{Moreggia}, D.~{Naumov},
  M.~{Pallavicini} et~al., \emph{{Full simulation of space-based extensive air
  showers detectors with ESAF}},
  \href{https://doi.org/10.1016/j.astropartphys.2010.02.005}{\emph{Astroparticle
  Physics} {\bfseries 33} (2010) 221}
  [\href{https://arxiv.org/abs/0907.5275}{{\ttfamily 0907.5275}}].

\bibitem{1995ApJ...441..144B}
D.J.~{Bird}, S.C.~{Corbato}, H.Y.~{Dai}, J.W.~{Elbert}, K.D.~{Green},
  M.A.~{Huang} et~al., \emph{{Detection of a Cosmic Ray with Measured Energy
  Well beyond the Expected Spectral Cutoff due to Cosmic Microwave Radiation}},
  \href{https://doi.org/10.1086/175344}{\emph{Astrophys.~J.} {\bfseries 441}
  (1995) 144} [\href{https://arxiv.org/abs/astro-ph/9410067}{{\ttfamily
  astro-ph/9410067}}].

\bibitem{2019ICRC...36..234D}
O.~{Deligny}, \emph{{The energy spectrum of ultra-high energy cosmic rays
  measured at the Pierre Auger Observatory and at the Telescope Array}},  in
  \emph{36th International Cosmic Ray Conference (ICRC2019)}, vol.~36 of
  \emph{International Cosmic Ray Conference}, p.~234, July, 2019.

\bibitem{TelescopeArray:2018xyi}
{\scshape Telescope Array} collaboration, \emph{{Depth of Ultra High Energy
  Cosmic Ray Induced Air Shower Maxima Measured by the Telescope Array Black
  Rock and Long Ridge FADC Fluorescence Detectors and Surface Array in Hybrid
  Mode}}, \href{https://doi.org/10.3847/1538-4357/aabad7}{\emph{Astrophys. J.}
  {\bfseries 858} (2018) 76}
  [\href{https://arxiv.org/abs/1801.09784}{{\ttfamily 1801.09784}}].

\bibitem{Spitzer1949}
L.~Spitzer, \emph{On the origin of heavy cosmic-ray particles},
  \href{https://doi.org/10.1103/PhysRev.76.583}{\emph{Phys. Rev.} {\bfseries
  76} (1949) 583}.

\bibitem{Hayakawa1972}
S.~Hayakawa, \emph{Dust grain origin of cosmic ray air showers},
  \href{https://doi.org/10.1007/BF00642736}{\emph{Astrophysics and Space
  Science} {\bfseries 16} (1972) 238}.

\bibitem{PhysRevD.61.087302}
L.A.~Anchordoqui, \emph{Cosmic dust grains strike again},
  \href{https://doi.org/10.1103/PhysRevD.61.087302}{\emph{Phys. Rev. D}
  {\bfseries 61} (2000) 087302}.

\bibitem{2001NuPhS..97..203A}
L.A.~{Anchordoqui}, M.T.~{Dova}, T.P.~{McCauley}, T.~{Paul}, S.~{Reucroft} and
  J.D.~{Swain}, \emph{{A pot of gold at the end of the cosmic ``raynbow''?}},
  \href{https://doi.org/10.1016/S0920-5632(01)01264-6}{\emph{Nuclear Physics B
  Proceedings Supplements} {\bfseries 97} (2001) 203}
  [\href{https://arxiv.org/abs/astro-ph/0006071}{{\ttfamily
  astro-ph/0006071}}].

\bibitem{1991aame.conf..434E}
N.N.~{Efimov} et~al., \emph{{Peculiarities of Muon Component in Giant EAS}},
  in \emph{Astrophysical Aspects of the Most Energetic Cosmic Rays},
  M.~{Nagano} and F.~{Takahara}, eds., p.~434, Jan., 1991.

\bibitem{PhysRevLett.56.1350}
T.W.~Atwater and P.S.~Freier, \emph{Meson multiplicity versus energy in
  relativistic nucleus-nucleus collisions},
  \href{https://doi.org/10.1103/PhysRevLett.56.1350}{\emph{Phys. Rev. Lett.}
  {\bfseries 56} (1986) 1350}.

\bibitem{2015ApJ...806..255H}
T.~{Hoang}, A.~{Lazarian} and R.~{Schlickeiser}, \emph{{On Origin and
  Destruction of Relativistic Dust and its Implication for Ultrahigh Energy
  Cosmic Rays}},
  \href{https://doi.org/10.1088/0004-637X/806/2/255}{\emph{Astrophysical
  Journal} {\bfseries 806} (2015) 255}
  [\href{https://arxiv.org/abs/1412.0578}{{\ttfamily 1412.0578}}].

\bibitem{Auger}
{\scshape Pierre Auger} collaboration, \emph{{Measurement of the cosmic-ray
  energy spectrum above $2.5{\times} 10^{18}$~eV using the Pierre Auger
  Observatory}}, \href{https://doi.org/10.1103/PhysRevD.102.062005}{\emph{Phys.
  Rev. D} {\bfseries 102} (2020) 062005}
  [\href{https://arxiv.org/abs/2008.06486}{{\ttfamily 2008.06486}}].

\bibitem{mini-EUSO}
S.~{Bacholle}, P.~{Barrillon}, M.~{Battisti}, A.~{Belov}, M.~{Bertaina},
  F.~{Bisconti} et~al., \emph{{Mini-EUSO} mission to study earth {UV} emissions
  on board the {ISS}},
  \href{https://doi.org/10.3847/1538-4365/abd93d}{\emph{The Astrophysical
  Journal Supplement Series} {\bfseries 253} (2021) 36}
  [\href{https://arxiv.org/abs/2010.01937}{{\ttfamily 2010.01937}}].

\bibitem{EUSO-SPB2}
J.~{Adams}, James~H., L.A.~{Anchordoqui}, J.A.~{Apple}, M.E.~{Bertaina},
  M.J.~{Christl}, F.~{Fenu} et~al., \emph{White paper on {EUSO-SPB2}},
  [\href{https://arxiv.org/abs/1703.04513}{{\ttfamily 1703.04513}}].

\bibitem{keuso}
F.~{Fenu}, S.~{Sharakin}, M.~{Zotov}, N.~{Sakaki} et~al., \emph{A performance study
  of the {K-EUSO} space-based observatory}, these proceedings, 2021.

\bibitem{POEMMA:2020ykm}
{\scshape POEMMA} collaboration, \emph{{The POEMMA (Probe of Extreme
  Multi-Messenger Astrophysics) observatory}},
  \href{https://doi.org/10.1088/1475-7516/2021/06/007}{\emph{JCAP} {\bfseries
  06} (2021) 007} [\href{https://arxiv.org/abs/2012.07945}{{\ttfamily
  2012.07945}}].

\end{thebibliography}
\end{document}